\documentclass[article,aps,showpacs,amsfonts,epsf]{revtex4}

\usepackage{graphicx}

\newcommand{\E}{{\mathcal{E}}}
\newcommand{\s}{\sigma}

\newcommand{\be}{\begin{equation}}
\newcommand{\ee}{\end{equation}}
\newcommand{\bea}{\begin{eqnarray}}
\newcommand{\eea}{\end{eqnarray}}
\newcommand{\ba}{\begin{array}}
\newcommand{\ea}{\end{array}}
\def\J#1#2#3#4{{#1} {\bf #2}, #3 (#4)}
\def\PRD{Phys. Rev. D}
\def\PR{Phys. Rev.}
\def\PRL{Phys. Rev. Lett.}
\def\PTP{Prog. Theor. Phys.}

\def\JMP{J. Math. Phys.}

\def\CQG{Class. Quantum Grav.}

\def\GRG{Gen. Relativ. Grav.}
\def\PLA{Phys. Lett. A}

\def\PTP{Prog. Theor. Phys.}

\def\PRSLA{Proc. R. Soc. Lond. A}

\def\NP{Nucl. Phys.}

\def\GC{{\it Gravit. Cosmology}}

\begin{document}
\draft
\title{Smarr formula for black holes endowed with both
electric and magnetic charges}

\author{V.~S.~Manko and H. Garc\'ia-Compe\'an}
\address{Departamento de F\'\i sica, Centro de Investigaci\'on y de
Estudios Avanzados del IPN, A.P. 14-740, 07000 M\'exico D.F.,
Mexico}

\begin{abstract}
The present paper clarifies how to consistently take into account
the contribution of magnetic charges in the generalized Smarr mass
formula by using properly the Tomimatsu's representation of Komar
integrals. It is shown in particular that in all three examples of
the dyonic solutions considered by us, the sum of the two
electromagnetic terms in Smarr's formula can be cast into the form
$\bar{\cal Q}\Phi^H_{ext}$, where ${\cal Q}$ is the complex charge
and $\Phi^H_{ext}$ the complex extension of the electric
potential.
\end{abstract}


\maketitle


\section{Introduction}

The well-known Smarr formula \cite{Sma}, the derivation of which
was inspired by the work of Christodoulou and Ruffini \cite{CRu},
relates concisely various physical characteristics of a single
black hole, and in its original version the black hole can only
carry the electric charge, being free of the magnetic one. Smarr's
mass formula and some of its possible generalizations were
extensively analyzed and discussed by Carter \cite{Car} with the
aid of the Komar integrals \cite{Kom}, and later Tomimatsu
\cite{Tom} succeeded in employing the Ernst formalism of complex
potentials \cite{Ern} for presenting Carter's general expressions
in a simple and elegant form suitable for the use even in the
multi-black-hole spacetimes. The specific problem studied by
Tomimatsu in \cite{Tom} was the application of the Ernst-Harrison
charging transformation \cite{Ern,Har} to the celebrated
double-Kerr solution of Kramer and Neugebauer \cite{KNe}, and in
particular he established that in that kind of binary systems the
usual Smarr formula was not satisfied generically by a charged
spinning black-hole constituent, what he associated with the
emergence of a Dirac string signalling about the presence of
magnetic charges. It should be emphasized that the correctness of
Tomimatsu's integral formulae was checked both analytically and
numerically in numerous papers devoted to binary black-hole
systems. Nonetheless, in the recent articles \cite{Cab1,Cab2}
dealing with a reparametrization of the Ernst-Manko-Ruiz (EMR)
solution for charged counterrotating sources \cite{EMR,SRo} some
modifications of the mass formula were implemented by the authors
that led them to physically inconsistent results, as has been
pointed out in \cite{MSa}. In this respect, it would be certainly
desirable, on the one hand, to figure out the precise source of
the problem lying behind the results of \cite{Cab1,Cab2} and, on
the other hand, to show how the correct use of Tomimatsu's
formulae is able to describe properly the electric and magnetic
charge contributions in the generalized Smarr formula. We also
note that the issue of the correct inclusion of the magnetic
charge parameter into exact solutions has an important
astrophysical dimension, in particular for modeling the exterior
fields of magnetized neutron stars, since the dipole magnetic
field may arise from the magnetic charges equal in absolute values
but opposite in sign.

In  order to make our discussion of the stationary dyonic dihole
case more comprehensible to the reader, in the next section we
will first consider the application of Tomimatsu's formulae to the
dyonic Kerr-Newman (KN) solution \cite{Car}. It will be seen that
this case does not represent any problems with regard to the
generalized Smarr formula and, moreover, can be helpful for
identifying the difficulty behind the modified mass formula of the
paper \cite{Cab2}. In Sec.~III the 5-parameter solution for a
stationary dyonic dihole will be reexamined with the aid of
Tomimatsu's formulae within the lines of the paper \cite{MRS}, and
in particular it will be explained why the Smarr formula in the
dyonic case may only slightly differ from the zero magnetic charge
case. A dyonic generalization of the Bret\'on-Manko (BM) solution
\cite{BMa,MRR} will be discussed in Sec.~IV as a non-trivial
extension of Sec.~III, with the idea to demonstrate possible
complications that may arise after the introduction of magnetic
charge into the solution. Concluding remarks are given in Sec.~V.

\section{The dyonic KN solution}

In this section we will show that in the case of the dyonic KN
solution \cite{Car}, which is a one-parameter generalization of
the usual KN metric \cite{NCC} and a four-parameter specialization
of the Demia\'nski-Newman electrovacuum spacetime \cite{DNe,McR},
the Tomimatsu's integrals are able to lead us straightforwardly to
the desired form of the generalized Smarr formula. Since these
integrals were originally designed for the work in the
Weyl-Papapetrou cylindrical coordinates $(\rho,z)$, it would be
likely to have a representation of the dyonic KN solution in the
latter coordinates. Note that this solution can be derived in a
rigorous way by means of Sibgatullin's integral method
\cite{Sib,MSi}, starting from the axis data
\be {\E}(\rho=0,z)=\frac{z-M-ia}{z+M-ia}, \quad
\Phi(\rho=0,z)=\frac{Q+i{\cal B}}{z+M-ia}, \label{KN_axis} \ee
where $M$, $a$, $Q$ and ${\cal B}$ are real parameters standing,
respectively, for the mass, angular momentum per unit mass,
electric and magnetic charges of the source. Then the resulting
Ernst potentials $\E$ and $\Phi$, as well as the metric functions
$f$, $\gamma$ and $\omega$ in the stationary axisymmetric line
element
\be d s^2=f^{-1}[e^{2\gamma}(d\rho^2+d z^2)+\rho^2d\varphi^2]-f(d
t-\omega d\varphi)^2, \label{pap} \ee
together with the electric $A_t$ and magnetic $A_\varphi$
components of the electromagnetic 4--potential have the form
\cite{AGM,MMR}
\bea \E&=&\frac{\s x-M-iay}{\s x+M-iay}, \quad
\Phi=\frac{Q+i{\cal B}}{\s x+M-iay}, \nonumber\\
f&=&\frac{\s^2(x^2-1)-a^2(1-y^2)}{(\s x+M)^2+a^2y^2}, \quad {\rm
e}^{2\gamma}=\frac{\s^2(x^2-1)-a^2(1-y^2)}{\s^2(x^2-y^2)},
\nonumber\\ \omega&=&-\frac{a(1-y^2)[2M(\s x+M)-Q^2-{\cal B}^2]}
{\s^2(x^2-1)-a^2(1-y^2)},\nonumber\\
A_t&=&-\frac{Q(\s
x+M)-a{\cal B}y}{(\s x+M)^2+a^2y^2}, \nonumber\\
A_\varphi&=&b_0-{\cal B}y+\frac{a(1-y^2)[Q(\s x+M)-a{\cal
B}y)]}{(\s x+M)^2+a^2y^2}, \nonumber\\
\s&=&\sqrt{M^2-a^2-Q^2-{\cal B}^2}, \label{EF_KN} \eea
where $(x,y)$ are related to $(\rho,z)$ by
\be x=\frac{1}{2\s}(r_++r_-), \quad y=\frac{1}{2\s}(r_+-r_-),
\quad r_\pm=\sqrt{\rho^2+(z\pm\s)^2}, \label{xy_rz} \ee
or, inversely,
\be \rho=\s\sqrt{(x^2-1)(1-y^2)}, \quad z=\s xy. \label{rz_xy} \ee

It should be noted that the expression for $A_\varphi$ in
(\ref{EF_KN}) contains the integration constant $b_0$ whose
particular choice, as will be seen later on, is very important for
obtaining a correct generalization of Smarr's mass formula.

The above formulas describe a non-extreme black hole when
$M^2>a^2+Q^2+{\cal B}^2$, and Tomimatsu's integrals for $M$,
$J=Ma$ and $Q$, supplemented with the formula for the magnetic
charge ${\cal B}$, have the form \cite{Tom}
\bea M&=&-\frac{1}{4}\int_{H}
\omega\Omega_{,z} d z, \label{kq1}\\
J&=&\frac{1}{4}\int_{H}\omega
\left[-1-{\textstyle\frac12}\omega\Omega_{,z}+\tilde A_\varphi
A'_{\varphi,z}+(A_\varphi A'_\varphi)_{,z}\right]
d z, \label{kq2}\\
Q&=&\frac{1}{2}\int_{H} \omega A'_{\varphi,z} d z, \label{kq3}\\
{\mathcal B}&=&\frac{1}{2}\int_{H} \omega A_{t,z} d z, \label{kq4}
\eea
with $\Omega={\rm Im}(\E)$, $A'_\varphi={\rm Im}(\Phi)$, $\tilde
A_\varphi=A_\varphi+\omega A_t$, all the functions entering
(\ref{kq1})-(\ref{kq4}) to be evaluated on the horizon which in
the $(\rho,z)$ coordinates is defined as the subset $\rho=0$,
$-\s<z<\s$ of the $z$-axis (see Fig.~1). There are also four more
black-hole characteristics involved in Tomimatsu's paper, which
are \cite{Car,Tom}
\be \kappa=\sqrt{-\omega^{-2}e^{-2\gamma}}, \quad
S=4\pi\sigma\sqrt{-\omega^{2}e^{2\gamma}}, \quad
\Omega^H=\omega^{-1}, \quad \Phi^H=-A_t-\Omega^H A_\varphi,
\label{kap} \ee
$\kappa$ being the surface gravity, $S$ the horizon's area,
$\Omega^H$ its angular velocity and $\Phi^H$ the electric
potential, and all of them are constant quantities because the
functions $\omega$, $\gamma$ and $\tilde A_\varphi$ take constant
values on the horizon.

We now turn to the mass formula of Smarr which (in the absence of
magnetic charge $\cal B$) reads
\be M=\frac{1}{4\pi}\kappa S+2J\Omega^H+Q\Phi^H
=\s+2J\Omega^H+Q\Phi^H. \label{Sma} \ee
Solving (\ref{Sma}) for $J$ and recalling that
$\omega=1/\Omega^H$, we get
\be J=\frac{\omega}{2}(-\s+M-Q\Phi^H), \label{J_form} \ee
whence it follows that it is equation (\ref{kq2}) for $J$ which
actually contains Smarr's formula: the correspondence of the first
two terms on the right hand sides of (\ref{J_form}) and
(\ref{kq2}) is trivially seen, while the correspondence between
the third terms is readily established if one takes into account
that $\tilde A_\varphi=-\omega\Phi^H$. For the usual KN solution,
the contribution of the fourth term on the right of (\ref{kq2}) is
zero if the constant $b_0$ in the potential $A_\varphi$ is set
equal to zero, which means that even in the absence of magnetic
charge the correct choice of $b_0$ is needed for the consistency
of Tomimatsu's formula (\ref{kq2}).

As was observed for instance in \cite{KKN}, Smarr's mass formula
in the presence of magnetic charge ${\cal B}$ is expected to have
the structure
\be M=\frac{1}{4\pi}\kappa S+2J\Omega^H+Q\Phi^H+{\cal B}\Phi^H_m,
\label{Sma_KNd} \ee
for taking into account the magnetic charge contribution via some
magnetic potential $\Phi^H_m$. Therefore it will be instructive to
work out the dyonic KN case in full detail using formulas
(\ref{EF_KN})-(\ref{kq4}). For this purpose we may start with the
evaluation of the potentials $\E$, $\Phi$, $A_t$, $A_\varphi$
(with $b_0=0$) and the metric functions (\ref{EF_KN}) on the
horizon where $r_\pm=\s\pm z$ and hence $x=1$, $y=z/\s$. The
resulting expressions for $\E$, $\Phi$, $A_t$ and $A_\varphi$ on
the horizon are
\bea \E&=&\frac{\s^2-M\s-iaz}{\s^2+M\s-iaz}, \quad
\Phi=\frac{\s(Q+i{\cal B})}{\s^2+M\s-iaz}, \nonumber\\
A_t&=&-\frac{Q\s^2(M+\s)-a{\cal B}\s z}{\s^2(M+\s)^2+a^2z^2},
\quad A_\varphi=-\frac{{\cal
B}z}{\s}+\frac{a(\s^2-z^2)[Q\s(M+\s)-a{\cal B}z]}
{\s[\s^2(M+\s)^2+a^2z^2]} \label{EF_hor} \eea
(the reader is reminded that the constant $b_0$ in the potential
$A_\varphi$ is set equal to zero), while both $\gamma$ and
$\omega$ on the horizon take constant values
\be e^{2\gamma}=-\frac{a^2}{\s^2}, \quad
\omega=\frac{(M+\s)^2+a^2}{a}. \label{go} \ee
Then from (\ref{EF_hor}) we get the form of $\Omega$ and
$A'_\varphi$,
\be \Omega={\rm Im}(\E)=-\frac{2Ma\s z}{\s^2(M+\s)^2+a^2z^2},
\quad A'_\varphi={\rm Im}(\Phi)=\frac{{\cal B}\s^2(M+\s)+aQ\s
z}{\s^2(M+\s)^2+a^2z^2}, \label{Om} \ee
and (\ref{EF_hor}) and (\ref{go}) permit us to see that the
combination $A_\varphi+\omega A_t$ is constant:
\be \tilde A_\varphi=-\frac{Q(M+\s)}{a}. \label{Ati} \ee

From (\ref{go}) and (\ref{kap}) we find the expressions for the
surface gravity, horizon's area and horizon's angular velocity:
\be \kappa=\frac{\s}{(M+\s)^2+a^2}, \quad S=4\pi[(M+\s)^2+a^2],
\quad \Omega^H=\frac{a}{(M+\s)^2+a^2}, \label{k_axis} \ee
while the form of the electric potential $\Phi^H$ is obtainable
from (\ref{kap}), (\ref{EF_hor}) and (\ref{k_axis}), yielding
\be \Phi^H=\frac{Q(M+\s)}{(M+\s)^2+a^2}. \label{FH} \ee

Since $\omega$ and $\tilde A_\varphi$ are constant quantities in
Tomimatsu's formulas (\ref{kq1})-(\ref{kq4}), the integration
there just reduces to evaluating the differences of the functions
at the points $z=+\s$ and $z=-\s$. Thus, for example, it is easy
to check that the relation (\ref{kq1}) for $M$ is satisfied
identically,
\be M=-\frac{1}{4}\omega[\Omega(z=+\s)-\Omega(z=-\s)]=M,
\label{M_id} \ee
and the same check can be readily performed in the formulas
(\ref{kq3}) and (\ref{kq4}) too. Obviously, the magnetic potential
$\Phi^H_m$ should be found from the Tomimatsu's formula
(\ref{kq2}) for $J$, and it is defined by the fourth term in the
integrand. Then, taking into account that
\be (A_\varphi A'_\varphi)_{z=+\s}-(A_\varphi A'_\varphi)_{z=-\s}
=-\frac{2{\cal B}^2(M+\s)}{(M+\s)^2+a^2}, \label{MAS} \ee
the formula for $J$ can be written as
\be J=\frac{\omega}{2}(-\s+M-Q\Phi^H-{\cal B}\Phi^H_m),
\label{J_KN} \ee
after the introduction of the magnetic potential $\Phi^H_m$ of the
form
\be \Phi^H_m=\frac{{\cal B}(M+\s)}{(M+\s)^2+a^2}. \label{Fm_KN}
\ee
Recalling that $\s=\kappa S/4\pi$ and $\omega=1/\Omega^H$, we
eventually arrive at the conclusion that the choice $b_0=0$ in the
potential $A_\varphi$ from (\ref{EF_KN}) ensures the verification
of the generalized Smarr formula by the KN dyon, Tomimatsu's
formula (\ref{kq2}) describing correctly the angular momentum
$J=Ma$ of the solution.

It is worth remarking that if the potential $A_\varphi$ had a
non-zero $b_0$, then the third and fourth terms on the right hand
side of (\ref{kq2}) would have modified equation (\ref{J_KN}) in
the following way:
\be J=\frac{\omega}{2}(-\s+M-Q\Phi^H-{\cal B}\Phi^H_m) +Qb_0,
\label{Jb0} \ee
thus violating the generalized Smarr formula (\ref{Sma_KNd}).
Precisely for that reason the constant $b_0$ in (\ref{EF_KN}) must
be set equal to zero; however, in some more complex dyonic
solutions the constant $b_0$, as will be seen in the next two
sections, must be assigned non-zero values to ensure consistent
verification of the mass formula (\ref{Sma_KNd}).

It should be also noted that the values of the potentials $\Phi^H$
and $\Phi^H_m$ of the KN dyon suggest that the last two
(electromagnetic) terms in the generalized Smarr formula
(\ref{Sma_KNd}) can be combined in one expression. Indeed, from
(\ref{FH}) and (\ref{Fm_KN}) we get
\be Q\Phi^H+{\cal B}\Phi^H_m=\frac{(Q^2+{\cal
B}^2)(M+\s)}{(M+\s)^2+a^2}, \label{Sem} \ee
so that by introducing the complex charge ${\cal Q}$ and the
extended electric potential $\Phi^H_{ext}$ via the formulas
\be {\cal Q}=Q+i{\cal B}, \quad \Phi^H_{ext}=\frac{{\cal
Q}(M+\s)}{(M+\s)^2+a^2}, \label{Fext} \ee
the left hand side of (\ref{Sem}) takes the form $\bar{\cal
Q}\Phi^H_{ext}$, and the generalized Smarr formula (\ref{Sma_KNd})
rewrites as
\be M=\frac{1}{4\pi}\kappa S+2J\Omega^H+\bar{\cal Q}\Phi^H_{ext},
\label{Sma_mod} \ee
thus being only slightly different from the usual mass formula
(\ref{Sma}). In what follows we shall see that the above form of
the generalized mass relation may also hold in the binary systems
of interacting dyons.

\section{The EMR dyonic dihole}

A more complicated dyonic model would be a pair of counterrotating
identical KN dyons whose electric (as well as magnetic) charges
are equal in absolute values but have opposite signs. Such model
is described by the EMR metric \cite{EMR} which contains as a
particular case the Emparan-Teo static electric dihole solution
\cite{ETe}), so that the more general model to be considered below
represents a stationary dyonic dihole with equatorial antisymmetry
\cite{EMR2}. A physical parametrization of the EMR solution
involving Komar quantities has been worked out in the paper
\cite{MRS}, the solution's potentials $\E$ and $\Phi$ in that
parametrization having the form
\bea \E&=&\frac{A-B}{A+B}, \quad \Phi=\frac{C}{A+B}, \nonumber\\
A&=&R^2(M^2-|{\mathcal Q}|^2\nu)(R_+-R_-)(r_+-r_-)+
4\s^2(M^2+|{\mathcal Q}|^2\nu)(R_+-r_+)(R_--r_-) \nonumber\\
&&+2R\s[R\s(R_+r_-+R_-r_+)+iMa\mu(R_+r_--R_-r_+)], \nonumber\\
B&=&2MR\s[R\s(R_++R_-+r_++r_-)-(2M^2-iMa\mu)(R_+-R_--r_++r_-)],
\nonumber\\
C&=&2C_0R\s[(R+2\s)(R\s-2M^2-iMa\mu)(r_+-R_-) +(R-2\s) \nonumber\\
&&\times(R\s+2M^2+iMa\mu)(r_--R_+)], \nonumber\\
R_\pm&=&\sqrt{\rho^2+(z+{\textstyle\frac{1}{2}}R\pm\sigma)^2},
\quad r_\pm=\sqrt{\rho^2+(z-{\textstyle\frac{1}{2}}R\pm\sigma)^2},
\label{EFn} \eea
where
\bea {\cal Q}&=&Q+i{\cal B}, \quad |{\mathcal Q}|^2=Q^2+{\cal
B}^2,
\nonumber\\
\s&=&\sqrt{M^2-\left(\frac{M^2a^2[(R+2M)^2+4|{\mathcal Q}|^2]}
{[M(R+2M)+|{\mathcal Q}|^2]^2}+|{\mathcal
Q}|^2\right)\frac{R-2M}{R+2M}}, \label{sigB} \eea
and $\mu$, $\nu$ and $C_0$ are dimensionless constant quantities
defined as
\bea &&\mu=\frac{R^2-4M^2}{M(R+2M)+|{\mathcal Q}|^2}, \quad
\nu=\frac{R^2-4M^2}{(R+2M)^2+4|{\mathcal Q}|^2}, \nonumber\\
&&C_0=-\frac{{\mathcal Q}(R^2-4M^2+2iMa\mu)}{(R+2M)(R^2-4\s^2)}.
\label{mnc} \eea
The real constants $M$, $Q$ and ${\cal B}$ are the mass, electric
and magnetic charges of the upper KN dyon, whose angular momentum
is $Ma$; the respective characteristics of the lower dyon are $M$,
$-Q$, $-{\cal B}$ and $-Ma$; $R$ is the separation coordinate
distance (see Fig.~2(a)). Remarkably, like in the case of a single
KN dyon, the charges $Q$ and ${\cal B}$ enter formulas
(\ref{EFn})-(\ref{mnc}) only in the combinations ${\cal
Q}=Q+i{\cal B}$ and ${\cal Q}\bar{\cal Q}\equiv|{\cal
Q}|^2=Q^2+{\cal B}^2$.

The corresponding metric coefficients $f$, $\gamma$ and $\omega$
are given by the expressions
\bea f&=&\frac{A\bar A-B\bar B+C\bar C}{(A+B)(\bar A+\bar B)},
\quad e^{2\gamma}=\frac{A\bar A-B\bar B+C\bar C}
{16R^4\s^4R_+R_-r_+r_-}, \quad \omega=-\frac{{\rm Im}[2G(\bar
A+\bar B)+C\bar I]}
{A\bar A-B\bar B+C\bar C}, \nonumber \\
G&=&-zB+R\s\{R(2M^2-|{\mathcal Q}|^2\nu)(R_-r_--R_+r_+)
+2\s(2M^2+|{\mathcal Q}|^2\nu)(r_+r_--R_+R_-)
\nonumber\\
&&+M[(R+2\s)(R\s-2M^2+iMa\mu)+2(R-2\s)|{\mathcal
Q}|^2\nu](R_+-r_-)
\nonumber\\
&&+M[(R-2\s)(R\s+2M^2-iMa\mu)-2(R+2\s)|{\mathcal
Q}|^2\nu](R_--r_+)\},
\nonumber\\
I&=&-zC+2C_0M[R^2(2M^2-2\s^2+iMa\mu)(R_+r_++R_-r_-)
\nonumber\\
&&+2\s^2(R^2-4M^2-2iMa\mu)(R_+R_-+r_+r_-)]-C_0(R^2-4\s^2) \nonumber\\
&&\times\{2M[R\s(R_+r_--R_-r_+)+(2M^2+iMa\mu)(R_+r_-+R_-r_+)] +R\s[R\s \nonumber\\
&&\times(R_++R_-+r_++r_-)+(6M^2+iMa\mu)(R_+-R_--r_++r_-)
+8MR\s]\}, \label{mfn} \eea
while the $t$ and $\varphi$ components of the electromagnetic
4-potential are given by the formulas
\be A_t=-{\rm Re}\left(\frac{C}{A+B}\right), \quad
A_\varphi=b_0+{\rm Im}\left(\frac{I}{A+B}\right), \label{tph} \ee
where $b_0$ is a real constant whose particular value should be
found from Tomimatsu's formula (\ref{kq2}).

Similar to the previous case considered in Sec.~II, it can be
checked that Tomimatsu's formulas (\ref{kq1}), (\ref{kq3}) and
(\ref{kq4}) are satisfied identically by the solution
(\ref{EFn})-(\ref{tph}) on both horizons, thus supporting the
interpretation of the parameters $M$, $Q$ and ${\cal B}$. At the
same time, computations performed in the formula (\ref{kq2}) for
$J$ on the upper horizon ($\rho=0$, ${\textstyle\frac12}R-\s\le
z\le {\textstyle\frac12}R+\s$) eventually lead to the equation
\be J=Ma+Q{\cal B}+b_0Q, \label{Jeq} \ee
whence we get
\be b_0=-{\cal B}, \label{boB} \ee
taking into account that $J=Ma$ by definition. On the other hand,
at the lower horizon ($\rho=0$, $-{\textstyle\frac12}R-\s\le z\le
-{\textstyle\frac12}R+\s$) the equation for the determination of
$b_0$ would take the form
\be -J=-Ma-Q{\cal B}-b_0Q, \label{Jeq2} \ee
thus being also consistent with the choice (\ref{boB}) for $b_0$.

Note that it is precisely the absence of the constant $b_0$ in the
potential $A_\varphi$ of the paper \cite{Cab2} that led the
authors of the latter to an unphysical redefinition $J-Q{\cal B}$
of the angular momentum throughout the solution, thus making their
results physically inconsistent (see \cite{MSa} for details).

The substitution $b_0=-{\cal B}$ into (\ref{tph}) ensures the
verification of the generalized Smarr formula (\ref{Sma_KNd}).
Indeed, the quantities $\kappa$, $S$, $\Omega^H$, $\Phi^H$ for the
upper dyon constituent can be shown to be determined by the
expressions
\bea \kappa&=&\frac{R\s[(R+2M)^2+4|{\mathcal Q}|^2]}
{(R+2M)^2[2(M+\s)(MR+2M^2+|{\mathcal Q}|^2)-|{\mathcal Q}|^2(R-2M)]}, \label{kS1}\\
S&=&\frac{4\pi}{R(R+2\s)}\left((R+2M)^2(M+\s)^2
+\frac{M^2a^2(R^2-4M^2)^2}{(MR+2M^2+|{\mathcal Q}|^2)^2}\right), \label{kS2}\\
\Omega^H&=&\frac{Ma[2(M-\s)(MR+2M^2+|{\mathcal Q}|^2)-|{\mathcal
Q}|^2(R-2M)]}
{(4M^2a^2+|{\mathcal Q}|^4)(MR+2M^2+|{\mathcal Q}|^2)}, \label{kS3}\\
\Phi^H&=&\frac{Q[|{\mathcal Q}|^2(M-\s)(MR+2M^2+|{\mathcal
Q}|^2)+2M^2a^2(R-2M)]} {(4M^2a^2+|{\mathcal
Q}|^4)(MR+2M^2+|{\mathcal Q}|^2)}, \label{kS4} \eea
while for the magnetic potential $\Phi^H_m$ the fourth term of
Tomimatsu's formula (\ref{kq2}) gives
\be \Phi^H_m={\mathcal B}\Phi^H/Q, \label{magp} \ee
and it is straightforward to check that (\ref{kS1})-(\ref{magp})
satisfy (\ref{Sma_KNd}) identically. Apparently, the lower dyon
constituent satisfies (\ref{Sma_KNd}) too, as the latter equation
is invariant under the sign change $J\to-J$,
$\Omega^H\to-\Omega^H$, $Q\to-Q$, $\Phi^H\to-\Phi^H$, ${\mathcal
B}\to-{\mathcal B}$, $\Phi^H_m\to-\Phi^H_m$.

It is also clear that $\Phi^H$ and $\Phi^H_m$ can be combined in
one potential of the form
\be \Phi^H_{ext}=\Phi^H+i\Phi^H_m= \frac{{\mathcal Q}[|{\mathcal
Q}|^2(M-\s)(MR+2M^2+|{\mathcal Q}|^2)+2M^2a^2(R-2M)]}
{(4M^2a^2+|{\mathcal Q}|^4)(MR+2M^2+|{\mathcal Q}|^2)},
\label{Fext2} \ee
in full analogy with the case of a sole KN dyon, and now the
quantities (\ref{kS1})-(\ref{kS3}) and (\ref{Fext2}) satisfy the
generalized Smarr formula in the form (\ref{Sma_mod}) which can
also be written as
\be M=\s+2J\Omega^H+\bar{\cal Q}\Phi^H_{ext}. \label{Sma_mod2} \ee
Moreover, it is not difficult to see that $\Phi^H_{ext}$ is
obtainable from the potential $\Phi^H$ of the ${\cal B}=0$
specialization of the general 5-parameter solution by formally
changing $Q$ to ${\cal Q}$ and $Q^2$ to $|{\mathcal Q}|^2$, so
that the generalized mass formula (\ref{Sma_mod}) could in
principle be viewed as obtainable from the usual Smarr formula
(\ref{Sma}) via the above parameter substitution, supplemented
with changing the resulting ${\cal Q}\Phi^H_{ext}$ to $\bar{\cal
Q}\Phi^H_{ext}$.

Since the magnetic lines of force of a magnetic dipole formed by a
pair of magnetic monopoles of opposite signs are somewhat
distinctive from those of a magnetic dipole generated by a
rotating electric charge, it would be interesting to illustrate
how the magnetic charge in the 5-parameter EMR solution affects
the magnetic lines of force of the 4-parameter stationary electric
dihole (${\cal B}=0$). In Fig.~3 all plots are defined by the same
values of $M$, $a$, $Q$ and $R$; however, the values of ${\cal B}$
are different. The magnetically uncharged case is depicted in
Fig.~3(a), and it is represented by two individual magnetic
dipoles generated by the KN black-hole constituents. In Fig.~3(b)
the dyonic constituents carry opposite magnetic charges that are
25 times less in magnitude than their electric charges, and the
magnetic lines of force are seen perturbed by the additional
magnetic field. By increasing twice the value of ${\cal B}$, the
magnetic lines of force in Fig.~3(c) become already qualitatively
those of two opposite magnetic charges.

\section{The dyonic BM solution}

In this section we shall consider an interesting example of a
binary dyonic system requiring a somewhat more subtle use of
Tomimatsu's formulae than in the case of a dyonic dihole.
Concretely, we are going to analyze a system composed of two
identical counterrotating KN dyons carrying the same
electric/magnetic charges (both in magnitude and signs). Such a
system is described by the solution obtainable from the original
BM metric \cite{BMa} in which the magnetic charge must be
introduced by means of the same parameter change $Q\to{\cal Q}$,
$Q^2\to|{\cal Q}|^2$ as in the case of the EMR 5-parameter
solution. The Ernst potentials $\E$ and $\Phi$ of the dyonic BM
solution thus have the form
\bea \E&=&\frac{A-B}{A+B}, \quad \Phi=\frac{C}{A+B}, \nonumber\\
A&=&(M^2-|{\mathcal
Q}|^2)[4\sigma^2(R_+R_-+r_+r_-)+R^2(R_+r_++R_-r_-)]
+[\sigma^2(R^2-4M^2+4|{\mathcal Q}|^2)  \nonumber\\
&&-M^2a^2R^2\mu](R_+r_-+R_-r_+) -2iaMR\mu\sigma(MR+2M^2-|{\mathcal
Q}|^2)(R_+r_--R_-r_+),
\nonumber\\
B&=&2MR\sigma\{R\sigma
(R_++R_-+r_++r_-)-[2(M^2-|{\mathcal Q}|^2)+iMa\mu(MR+2M^2-|{\mathcal Q}|^2)] \nonumber\\
&&\times(R_+-R_--r_++r_-)\},
\nonumber\\
C&=&{\cal Q}B/M, \nonumber\\
R_\pm&=&\sqrt{\rho^2+(z+\frac{1}{2}R\pm\sigma)^2}, \quad
r_\pm=\sqrt{\rho^2+(z-\frac{1}{2}R\pm\sigma)^2}, \label{EP} \eea
where
\bea {\cal Q}&=&Q+i{\cal B}, \quad |{\mathcal Q}|^2=Q^2+{\cal
B}^2,
\nonumber\\
\sigma&=&\sqrt{M^2-|{\mathcal Q}|^2-M^2a^2\mu}, \quad
\mu=\frac{R^2-4M^2+4|{\mathcal Q}|^2}{(MR+2M^2-|{\mathcal
Q}|^2)^2}, \label{sigBM} \eea
and the parameters $M$, $-a$, $Q$, ${\cal B}$ stand for the mass,
angular momentum per unit mass, electric and magnetic charges of
the upper dyon constituent, while all the characteristics of the
lower dyon are the same, except that its angular momentum per unit
mass has opposite sign, $+a$ (see Fig.~2(b)).

The metric functions $f$, $\gamma$ and $\omega$ of the dyonic BM
solution, together with the potentials $A_t$ and $A_\varphi$
describing the electromagnetic field, have the form
\medskip
\bea f&=&\frac{A\bar A-B\bar B+C\bar C}{(A+B)(\bar A+\bar B)},
\quad e^{2\gamma}=\frac{A\bar A-B\bar B+C\bar
C}{16R^4\sigma^4R_+R_-r_+r_-}, \quad \omega=-\frac{{\rm
Im}[2G(\bar A+\bar B)+C\bar I]}{A\bar A-B\bar B+C\bar C},
\nonumber\\
G&=&-zB+R\sigma\{(2M^2-|{\mathcal
Q}|^2)[2\sigma(r_+r_--R_+R_-)+R(R_-r_--R_+r_+)]
\nonumber\\
&&+M(R+2\sigma)[R\sigma-2(M^2-|{\mathcal Q}|^2)-
iMa\mu(MR+2M^2-|{\mathcal Q}|^2)](R_+-r_-) \nonumber\\
&&+M(R-2\sigma)[R\sigma+2(M^2-|{\mathcal
Q}|^2)+iMa\mu(MR+2M^2-|{\mathcal Q}|^2)](R_--r_+)\},
\nonumber\\
I&=&\frac{{\cal Q}}{M}\{G+R|{\mathcal Q}|^2
\sigma[2\sigma(r_+r_--R_+R_-)+R(R_-r_--R_+r_+)]\}, \nonumber\\
A_t&=&-{\rm Re}\left(\frac{C}{A+B}\right), \quad
A_\varphi=b_0+{\rm Im}\left(\frac{I}{A+B}\right),
\label{metric_BM} \eea
where we again introduced the arbitrary constant $b_0$ in the
expression of $A_\varphi$.

On the upper and lower horizons (${\textstyle\frac12}R-\s\le z\le
{\textstyle\frac12}R+\s$ and $-{\textstyle\frac12}R-\s\le z\le
-{\textstyle\frac12}R+\s$ parts of the $z$-axis, respectively),
Tomimatsu's formulas (\ref{kq1}), (\ref{kq3}) and (\ref{kq4}) are
satisfied identically, thus confirming the interpretation of the
parameters $M$, $Q$ and ${\cal B}$ as the mass, electric charge
and magnetic charge of each dyon constituent. As for the angular
momentum of the upper dyon defined as $-J=-Ma$, Tomimatsu's
formula (\ref{kq2}) leads instead of an identity to the equation
\be -J=-Ma-Q{\cal B}+b_0Q, \label{Jueq} \ee
whence it follows immediately that the constant $b_0$ must have
the form
\be b_0={\cal B} \label{b0u} \ee
in order to satisfy (\ref{kq2}) identically on the upper horizon.
Now, making use of the results of the paper \cite{MRS}, it is not
difficult to show that the quantities $\kappa$, $S$, $\Omega^H$
and $\Phi^H$ of the upper horizon are given by the expressions
\bea \kappa&=&\frac{R\s}{\Delta}, \quad S=\frac{4\pi\Delta}{R},
\quad \Omega^H=-\frac{Ma\mu(MR+2M^2-|{\mathcal Q}|^2)}{\Delta}, \nonumber\\
\Phi^H&=&\frac{Q[(R+2M)(M+\s)-2|{\mathcal Q}|^2]}{\Delta}, \nonumber\\
\Delta&=&2M(R+2M)(M+\s)-|{\mathcal Q}|^2(R+4M+2\s),
\label{kap_res} \eea
and to these we should aggregate the expression of the magnetic
potential $\Phi^H_m$ corresponding to the fourth term in the
formula (\ref{kq2}):
\be \Phi^H_m=\frac{{\cal B}[(R+2M)(M+\s)-2|{\mathcal
Q}|^2]}{\Delta}. \label{FmBM} \ee
Then a simple check indicates that the quantities (\ref{kap_res})
and (\ref{FmBM}) verify the generalized Smarr formula
(\ref{Sma_KNd}) or, by introducing
\be \Phi^H_{ext}=\frac{{\cal Q}[(R+2M)(M+\s)-2|{\mathcal
Q}|^2]}{\Delta}, \label{FextBM} \ee
the formula (\ref{Sma_mod}).

At this point, turning to the lower horizon, it might seem
plausible to make use of the symmetry of the dyonic configuration
under consideration and just conclude that the lower dyon verifies
the generalized mass formula (\ref{Sma_KNd}) too as it has the
same physical characteristics (\ref{kap_res}) and (\ref{FmBM}) as
the upper one (albeit the sign change in $\Omega^H$ that does not
affect (\ref{Sma_KNd}) because $J$ also changes its sign).
However, in reality, coming to such an obvious physical conclusion
is not straightforward at all, since Tomimatsu's formula
(\ref{kq2}) leads on the lower horizon to the equation
\be J=Ma+Q{\cal B}+b_0Q, \label{Jleq} \ee
which becomes an identity only at
\be b_0=-{\cal B}, \label{b0l} \ee
this value of $b_0$ being different from the one in (\ref{b0u}).
This difference, which was absent in the previous case of the
dyonic EMR solution, is explained by the fact that the term $b_0Q$
on the right hand sides of (\ref{Jueq}) and (\ref{Jleq}) has the
same sign since the BM dyons carry identical charges, while the
charges of the EMR dyons are opposite in sign and hence the term
$b_0Q$ enters equations (\ref{Jeq}) and (\ref{Jeq2}) with
different signs, thus not causing problems of consistency.

To resolve the above (actually spurious) problem of two-valued
$b_0$, we have to resort to the help of the well-known papers of
Wu and Yang \cite{WYa1,WYa2} on the Dirac monopole \cite{Dir}
where for disposing of the Dirac string singularity two regions
were used to define a pair of potentials $A_\varphi$, differing on
the intersection of these regions by a gauge transformation (a
more general system of various interacting magnetic monopoles,
electrons and photons has been considered in \cite{Wan} within the
same basic concept). Wu and Yang's idea of exploiting the gauge
freedom for adjusting appropriately the magnetic component
$A_\varphi$ of the electromagnetic 4-potential was implemented in
the general relativistic theory by Semiz \cite{Sem} who introduced
for his specific purposes a double-valued constant corresponding
to two different regions precisely in the context of the dyonic KN
solution. Following the aforementioned papers, we shall define
$A_\varphi$ of the dyonic BM solution on two domains, the first
one with its gauge determined by the value $b_0={\cal B}$ for
$z\ge0$, and the other one with the second gauge defined by the
value $b_0=-{\cal B}$ for $z<0$. Such a redefinition of
$A_\varphi$ is not only likely but in fact required in order for
the potential $A_\varphi$ to be consistent with the symmetry of
our dyonic configuration after the choice $b_0={\cal B}$ has been
made on the upper horizon. Therefore, the constant $b_0$ in
(\ref{metric_BM}) must be finally assigned the double value
\be b_0=\pm{\cal B}, \label{b0f} \ee
where the plus and minus signs correspond to the two domains used
to define the potential $A_\varphi$. As a consequence, Tomimatsu's
formula (\ref{kq2}) works perfectly well on both horizons of the
dyonic BM solution where the generalized mass relations
(\ref{Sma_KNd}) and (\ref{Sma_mod}) are verified identically.

\section{Conclusions}

In the present paper we have shown that Tomimatsu's integral
formulae describe consistently the contribution of magnetic charge
in the generalized Smarr mass relation. The lack of physical
consistency of the results in \cite{Cab1,Cab2} is therefore
explained exclusively by incautious use of the expression for the
potential $A_\varphi$ in Tomimatsu's formula defining the angular
momentum, which consists in overlooking the fact that $A_\varphi$
must contain an arbitrary additive constant $b_0$ whose correct
choice is vital for all the physical interpretations. It is
interesting that this constant must be chosen distinctly in each
of the three dyonic solutions considered in our paper, which may
lead us to the following conclusions. First, adding the magnetic
charge to a single KN black hole does not really complicate the
use of Tomimatsu's formulae, and consequently the extension of
Smarr's mass relation, because $b_0$ must be assigned the same
zero value as in the case of the magnetically uncharged KN
solution. Second, in the dyonic dihole case the constant $b_0$ is
compelled to take a non-zero value in order to cancel out the term
$Q{\cal B}$ in Tomimatsu's formula (\ref{kq2}) that arises due to
the interaction of dyons; it is clear as well that the above term
requires the presence of both electric and magnetic charges, so
that in the absence of one of these the constant $b_0$ becomes
zero. Third, the dyonic BM solution has revealed that the constant
$b_0$ can also be a multi-valued quantity depending on the number
of domains used to determine the potential $A_\varphi$, and
Tomimatsu's formulae play an important role in finding the
particular values of such $b_0$. This example actually confirms
the words of Cheng and Li that ``the Dirac string is a gauge
artefact'' \cite{CLi}.

It is remarkable that in all our examples of dyonic spacetimes the
generalized mass formula takes a very simple form (\ref{Sma_mod})
only slightly different from the usual Smarr's formula. This is a
consequence of the relation $\Phi^H/Q=\Phi^H_m/{\cal B}$ existing
between the potentials $\Phi^H$ and $\Phi^H_m$ in those particular
examples, and it would be interesting to see in the future how
generically the above relation holds in the multi-dyonic
configurations possessing less symmetry.

\section*{Acknowledgments}

This paper is dedicated to the memory of Jos\'e Antonio Aguilar
S\'anchez. We are grateful to the referees for valuable remarks
and literature suggestions that helped to improve the
presentation. This work was partially supported by Project~128761
from CONACyT of Mexico.

\newpage

\begin{figure}
  \centering
    \includegraphics[width=8cm]{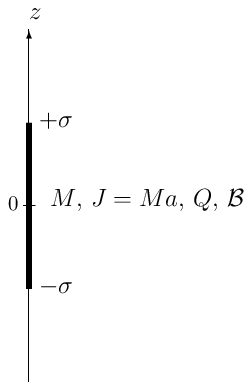}
  \caption{Location
of the KN dyon on the symmetry axis.}
  \label{fig1}
\end{figure}

\begin{figure}
  \centering
    \includegraphics[width=12cm]{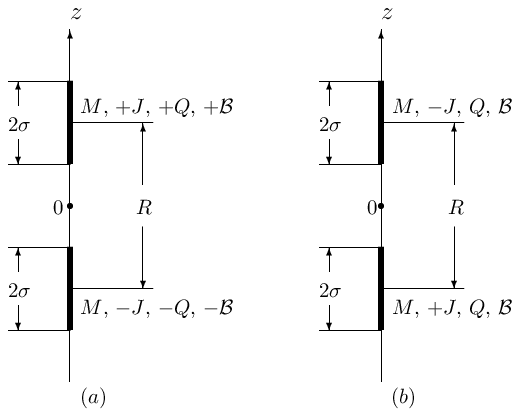}
  \caption{Location
of the dyonic black holes on the symmetry axis (a) in the case of
the EMR dyonic solution, and (b) in the case of the dyonic BM
solution.}
  \label{fig1}
\end{figure}

\begin{figure}
  \centering
    \includegraphics[width=16cm]{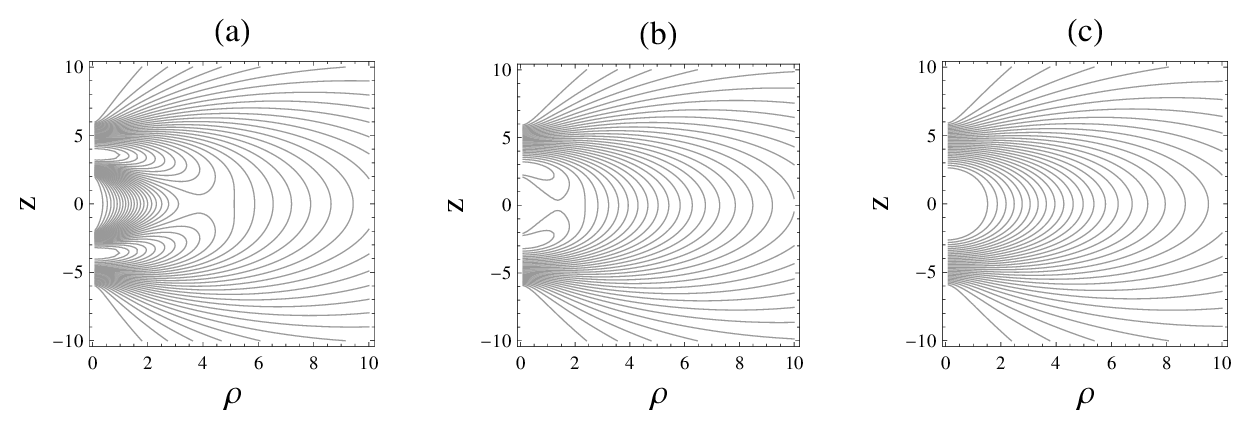}
\caption{Magnetic lines of force in the case of the EMR solution.
The particular values of the parameters are the following: $M=2$,
$a=0.25$, $Q=0.5$, $R=8$ (for all plots), and ${\cal B}=0$,
$0.02$, $0.04$ for (a), (b) and (c), respectively.}
  \label{fig1}
\end{figure}

\end{document}